\documentclass[9pt,twocolumn,twoside]{osajnl}

\journal{ol}

\setboolean{shortarticle}{false}

\usepackage{mathptmx}
\usepackage{xcolor}
\usepackage{todonotes}
\presetkeys{todonotes}{inline,inlinewidth=4cm}{}
\definecolor{linkblue}{RGB}{31,119,180}
\usepackage{siunitx}
\usepackage{graphicx}
\usepackage{amsmath}
\usepackage{amssymb}
\usepackage{natbib}
\usepackage{textcomp}
\usepackage{listings}
\usepackage{courier}

\usepackage{booktabs}
\usepackage{pgf}

\newcommand{\gsw}{gravitational-wave}
\newcommand{\chiEff}{$d_\mathrm{Q}=\SI{4.75(18)}{pm/V}$}
\newcommand{\efficiency}{$\eta=\SI{88.3(14)}\%$}

\newcommand{\visibility}{$V = 0.975$}

\begin{document}

\title{Highly-efficient generation of coherent light at 2128\,nm via degenerate optical-parametric oscillation}
\author[1]{Christian Darsow-Fromm}
\author[1]{Maik Schröder}
\author[1,2]{Julian Gurs}
\author[1]{Roman Schnabel}
\author[3,4]{Sebastian Steinlechner}
\affil[1]{Institut für Laserphysik und Zentrum für Optische Quantentechnologien der Universität Hamburg, Luruper Chaussee 149, 22761 Hamburg, Germany}
\affil[2]{3rd Institute of Physics, University of Stuttgart and Institute for Quantum Science and Technology IQST, 70569, Stuttgart, Germany}
\affil[3]{Faculty of Science and Engineering, Maastricht University, Duboisdomein 30, 6229 GT Maastricht, The Netherlands}
\affil[4]{Nikhef, Science Park 105, 1098 XG Amsterdam, The Netherlands}

\begin{abstract}
 Cryogenic operation in conjunction with new test-mass materials promises to reduce the sensitivity limitations from thermal noise in gravitational-wave detectors.
The currently most advanced materials under discussion are crystalline silicon as a substrate with amorphous silicon-based coatings. They require, however, operational wavelengths around 2\,\textmu m to avoid laser absorption.
Here, we present a light source at 2128\,nm based on a degenerate optical parametric oscillator (DOPO) to convert light from a 1064\,nm non-planar ring-oscillator (NPRO).
We achieve an external conversion efficiency of (88.3\,±\,1.4)\,\% at a pump power of 52\,mW in PPKTP (periodically-poled potassium titanyl phosphate, internal efficiency was 94\,\%), from which we infer an effective non-linearity of (4.75\,±\,0.18)\,pm/V.
With our approach, light from the established and existing laser sources can be efficiently converted to the 2\,\textmu m regime, while retaining the excellent stability properties.

\end{abstract}

\maketitle

\section{Introduction}
\label{sec:introduction}

Since the observation of the first black-hole coalescence in 2015, \gsw\ detection has evolved from proof-of-principle experiments into the new field of \gsw\ astronomy \cite{abbottObservationGravitationalWaves2016,ligoscientificcollaborationandvirgocollaborationGW170817ObservationGravitational2017,abbottGW190814GravitationalWaves2020}.
Further increasing the sensitivity of detectors not only allows the observation of weaker signals, but will also expand the detection range towards the entire  universe.
This promises new insights into cosmology and even the origin of the universe by statistical evaluation of the \gsw\ background \cite{shoemakerGravitationalWaveAstronomy2020s2019}. 
Especially the low-frequency regime is of interest in the context of multi-messenger astronomy, since merger events cross the \gsw\ spectrum days to weeks before the coalescence, giving ample pre-warning for a precise sky localization of any electro-magnetic counterpart.

Enhancement of the detector sensitivity, however, is a highly complex task involving many fields of expertise.
For instance, coating thermal noise poses a significant limitation of current detectors, particularly in the mid-frequency range from several tens of Hz to a few hundreds of Hz \cite{levinInternalThermalNoise1998,harryThermalNoiseInterferometric2002a,barsottiUpdatedAdvancedLIGO2018}.
This will be solved by operation at cryogenic temperatures and a simultaneous change of the mirror substrate and coating materials.
The most promising substrate candidate, crystalline silicon, excells with high mechanical Q-factor and thermal conductivity in the cryogenic regime, in contrast to the currently used fused silica \cite{hildXylophoneConfigurationThirdgeneration2010,etscienceteamEinsteinGravitationalWave2011}.
Latest research in coating technology, on the other hand, has shown promising mechanical loss results with amorphous silicon thin films \cite{murrayIonbeamSputteredAmorphous2015}.
Optimized coatings utilizing this material could show a thermal noise that is a factor of twelve below conventional silica-tantala coatings \cite{steinlechnerSiliconBasedOpticalMirror2018a}.
However, the operation wavelength with these novel coatings is restricted to above \SI{1.8}{\micro m}.
Otherwise the absorption would exceed the required order of a few ppm \cite{etscienceteamEinsteinGravitationalWave2011,ligoscientificcollaborationInstrumentScienceWhite2019,greenOpticalPropertiesIntrinsic1995}, which would lead to significant heating of the test masses, distortions from thermal lensing and hinder advanced quantum sensing techniques.
As such, design studies for upcoming detector generations such as LIGO Voyager and Cosmic Explorer \cite{ligoscientificcollaborationInstrumentScienceWhite2019} feature wavelengths around \SI{2}{\micro m}. Prototype facilities like ETpathfinder \cite{ETpathfinderDesignReport2020} are planning to investigate interferometry with this novel wavelength for \gsw{} detection.

New laser schemes have to provide a comprehensive solution for high precision quantum metrology, including optics and detection devices \cite{Schnabel2010,LSC2011,Tse2019,Acernese2019}.
In addition, the laser sources themselves have stringent requirements in terms of power stability, amplitude and phase noise, as well as spatial mode quality.
It took decades of development effort to reach the technological matureness with laser sources at \SI{1064}{nm}.
Laser technologies at around \SI{2}{\micro m} are most often based on either holmium- or thulium-doped laser media, but their maturity and performance levels still leave a lot to be desired \cite{mansellObservationSqueezedLight2018}, as their conventional use in medical and LIDAR applications does not require exceptional stability.
In contrast, our approach employs laser sources at \SI{1064}{nm} currently used in gravitational-wave detection and converts the light via optical parametric down-conversion to \SI{2128}{nm}.
This nonlinear process is known to retain the stability and noise properties of the pump \cite{eckardtOpticalParametricOscillator1991a,naborsCoherencePropertiesDoubly1990}, allowing us to fully take advantage of the already optimized technology.

Prior optical parametric oscillators have been used e.g.\ to provide tunable continuous-wave laser sources
\cite{wongHighconversionefficiencyWidelytunableAllfiber2007a}, 
wavelength conversion of pumped lasers
\cite{arisholmOpticalParametricMaster2004}, 
or even to generate quantum random bits, exploiting their inherent bistability
\cite{marandiAllopticalQuantumRandom2012a}. 

Here we report highly efficient generation of coherent light at \SI{2128}{nm} employing an external degenerate optical parametric oscillator (DOPO) and verify the preservation of power stability, amplitude and phase noise, as well as spatial mode quality.

\section{Experimental Setup}
\begin{figure}[b!]
    \includegraphics[width=\linewidth]{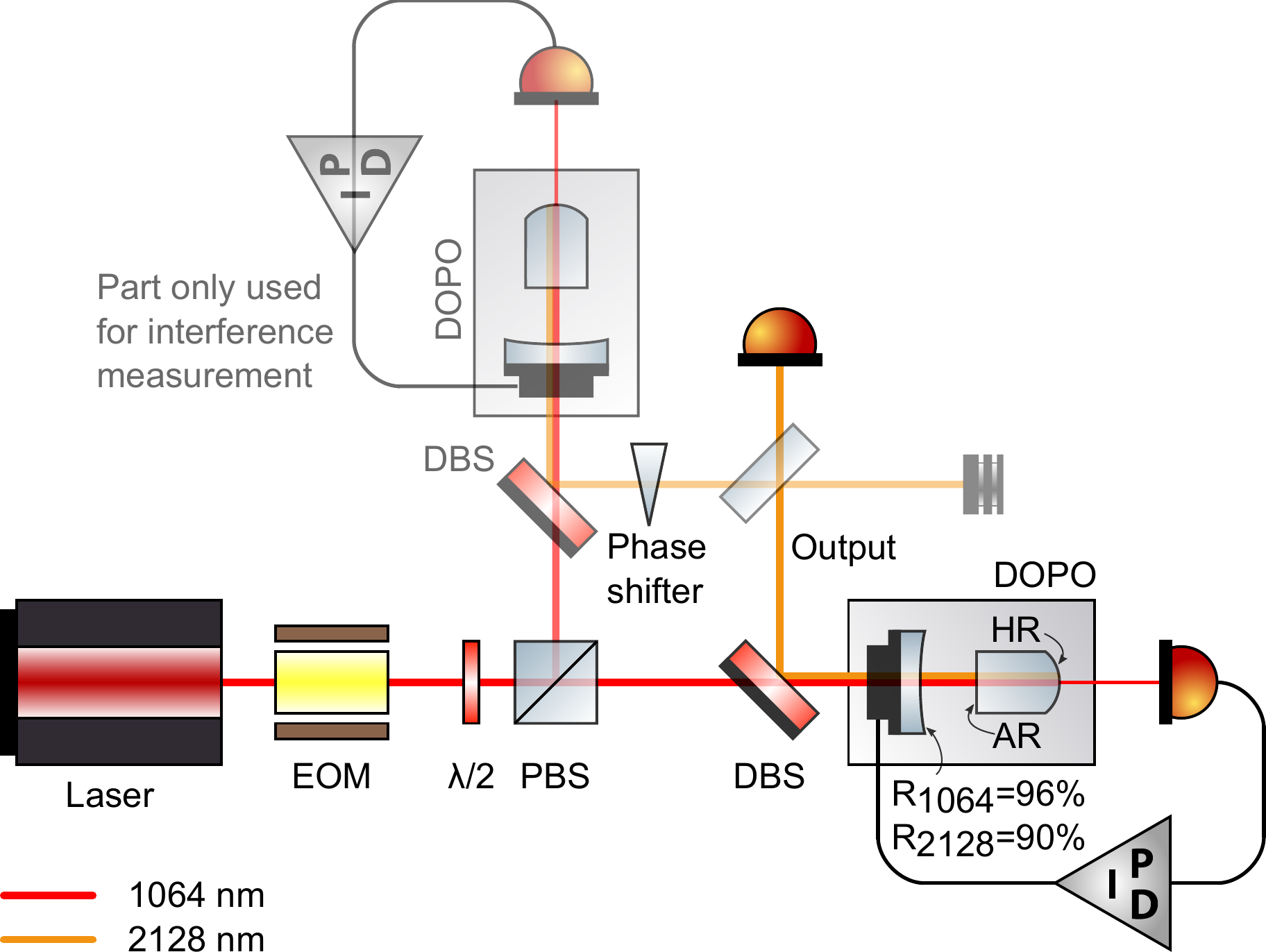}
    \caption{
        Simplified schematics of our experiment.
        The laser was a NPRO with \SI{2}{W} output power at the wavelength $\lambda = \SI{1064}{nm}$.
        An electro-optical modulator (EOM) provided phase-modulation sidebands at \SI{28}{MHz} for a Pound-Drever-Hall locking scheme of the cavity.
        A combination of half-waveplate and polarizing beam-slitter (PBS) adjusted the pump power.
        The converted light was split from the pumping beam with a dichroic beam-splitter (DBS).
        For diagnostic purposes, the converted light could be send towards a spectrometer and confocal cavity.
        A second, identical DOPO was installed together with a phase shifter and 50/50 beam-splitter for the interference measurement (sec.~\ref{sec:interference}).
    }
    \label{fig:lasersystem}
\end{figure}

Our setup for the generation of \SI{2128}{nm} light was based on type 0 degenerate optical-parametric oscillation (DOPO) in a nonlinear resonator, pumped by \SI{1064}{nm} light from an nonplanar ring oscillator (NPRO)~\cite{kaneMonolithicUnidirectionalSinglemode1985} master laser, see figure~\ref{fig:lasersystem}.
We chose a hemilithic resonator design around a periodically-poled potassium titanyl phosphate (PPKTP) crystal, where one side of the crystal functioned as end mirror of the resonator.
This side was highly reflective for both wavelengths, while the other side of the crystal was anti-reflective.
The coupling mirror was mounted on a piezo to scan and stabilize the cavity length and was coated with a reflectivity of 96\,\% at \SI{1064}{nm} and 90\,\% at \SI{2128}{nm}.
The overall cavity length has been simulated and optimized to suppress higher-order Gaussian modes.
Further optical properties of the DOPO are summarized in table~\ref{tab:cavity}.
Our cavity was coupled with a mode-matching efficiency of \SI{94}{\%}, which was mainly limited by the beam shape of the pump laser.
We used a modified Pound-Drever-Hall control scheme in transmission of the resonators together with a digital controller \cite{darsow-frommNQontrolOpensourcePlatform2020} to stabilize the DOPO cavity.

To simultaneously achieve the degeneracy of signal and idler fields, as well as a high conversion efficiency, we precisely controlled the temperature of two regions of the crystal \cite{zielinskaFullyresonantTunableMonolithic2017,schonbeckaxelCompactSqueezedlightSource2018}.
The main temperature $T_1$ adjusted the quasi phase-matching for the degenerate process.
About \SI{1}{mm} of the crystal's high-reflective (HR) end was separately temperature controlled ($T_2$) to ensure resonance for \SI{1064}{nm} and \SI{2128}{nm} at the same time.

\begin{table}[t]
    \centering
    \begin{tabular}{lrrl}
        \hline
                & 1064         &  2128 & nm \\
        \hline
        finesse & 153   &  59.5   & \\
        waist radius  & 33.2  &  47.4   & \textmu m \\
        free spectral range & 3.80  &  3.83   & GHz \\
        linewidth (FWHM) & 24.9  &  64.3   & MHz \\
        coupler reflectivity & 96 & 90 & \% \\
        \hline
    \end{tabular}
    \caption{Optical properties of the DOPO cavity for the pump and converted wavelengths.}
    \label{tab:cavity}
\end{table}

The \SI{1064}{nm} and \SI{2128}{nm} fields were separated from each other by a dichroic beam splitter and the power of the converted light was tracked by a photodiode. 
To determine the wavelengths of signal and idler fields, we used a Bruker Equinox 55 spectrometer with a resolution of \SI{0.5}{cm^{-1}}, which corresponds to \SI{0.23}{nm} at $\lambda=\SI{2128}{nm}$.

DOPOs are known to be highly sensitive to reflection of light back into the resonator
\cite{falkInstabilitiesDoublyResonant1971}, 
which is caused by a bistability allowing the randomly chosen phase states $0$ and $\pi$ for the converted light
\cite{naborsCoherencePropertiesDoubly1990,
marandiAllopticalQuantumRandom2012a}. 
For this reason, we tried to avoid back-reflections as much as possible by slightly tilting all optics after the DOPO.
In addition, we found an operation above the point of maximum efficiency to be much less sensitive than below this point.
An output power of about \SI{80}{mW} has been shown to be stable for our configuration.

\section{Results/Characterization}
\label{sec:results}
\begin{figure}[b!]
    \resizebox{\linewidth}{!}{\input{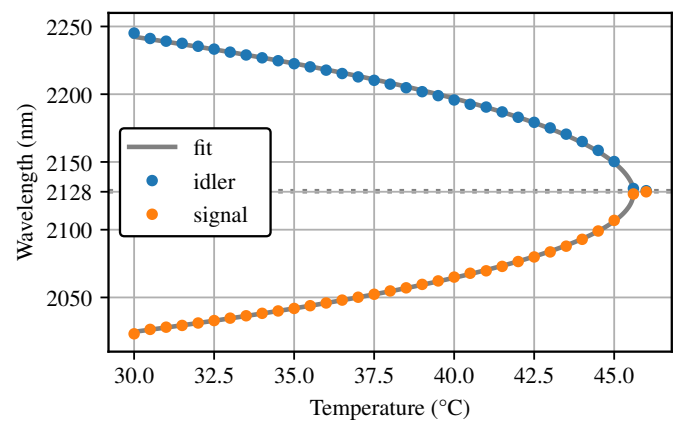}}
    \caption{
        Measured signal and idler wavelengths versus crystal temperature ($T_1 = T_2$).
        Below the temperature of degeneracy (about \SI{45.1}{\degree C} for this crystal) the difference of the signal and idler wavelengths increased with decreasing temperature.
        In this region the power of the converted light was nearly constant.
        At degeneracy the output power dropped quickly when the temperature was further increased and the conversion stopped completely at temperatures exceeding \SI{46}{\degree C}.
    }
    \label{fig:lambda_T}
\end{figure}

We characterized our setup to validate the performance required for interferometric light sources in \gsw\ detection.
This includes temperature tuning behavior of the oscillation wavelength, conversion efficiency, interferometric visibility, and power stability.
We would have liked to also include an amplitude noise spectrum of the DOPO output field, but were unable to obtain a sufficiently broadband, low-noise and high dynamic range photo detector for \SI{2}{\micro m}.

\subsection{Wavelength dependency on the crystal temperature}
\label{sec:wavelength_dependency}

Tuning the crystal temperature $T_1$ varies the wavelengths of the signal and idler beams, as it influences the phase-matching condition of the nonlinear crystal.
The main effect results from the temperature-dependent refractive index, while the thermal expansion has a relatively small contribution \cite{smithThermoopticThermalExpansion2016b}.
Figure~\ref{fig:lambda_T} shows the wavelength dependency of signal and idler on the temperature, measured with the spectrometer.
Our measurement shows the expected temperature dependency near degeneracy,
$ \Delta\omega = k \sqrt{T - T_0} $
\cite{giordmaineTunableCoherentParametric1965}, where $\Delta\omega$ is the difference in frequency between signal and idler.
Only in a small temperature region degeneracy could be reached, requiring a fine tuning of the phase-matching temperature.

When increasing the crystal temperature, the output power of the converted light stays roughly constant up to the point of degeneracy.
Above this point, the output power quickly dropped as energy conservation and phase matching condition could no longer be satisfied simultaneously.

\subsection{Conversion efficiency}

\begin{figure}[t]
    \resizebox{\linewidth}{!}{\input{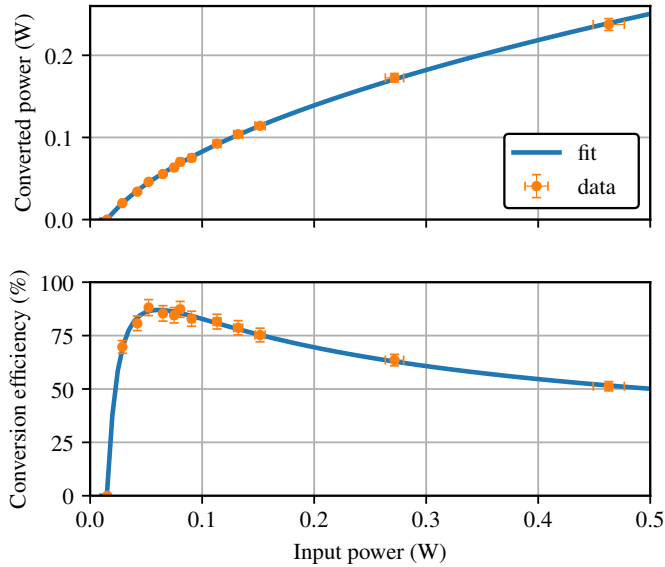}}
    \caption{
        Power of the converted light (top) and external conversion efficiency (bottom). The indicated error bars correspond to the measurement accuracy of the thermal power meter of \SI 3\%.
        The blue lines show our best fit to the data from a time-domain simulation model.
        At the point of maximum conversion efficiency, \SI{52}{mW} of pump power was converted into \SI{46}{mW} of output power.}
    \label{fig:conversion_efficiency}
\end{figure}

The conversion efficiency $\eta$ depending on the pump power was measured by inserting a thermal power head into the pump and output fields, respectively.
Figure~\ref{fig:conversion_efficiency} shows the measured power levels and the calculated external conversion efficiency, i.e.\ not corrected for power loss from mode-matching, reflection loss of the crystal's AR coating, internal absorption, as well as residual transmission through the resonator.
We fitted a numerical model to the measured data with the time-domain simulation program NLCS \cite{lastzkaNumericalModellingClassical2010}.
As free parameters for the simulation, we used the effective non-linearity $d_Q$ and maximum external conversion efficiency $\eta$.
Here, $d_Q$ is related to the nonlinear coefficient $d_{33}$ of our crystal geometry by an additional Fourier factor introduced by the quasi phase-matching, $d_Q = \frac 2\pi d_{33}$.

We obtained \chiEff{} for the harmonic transition from \SI{1064}{nm} to \SI{2128}{nm} in quasi phase-matched PPKTP.
The result for the maximum external conversion efficiency was \efficiency{} at an incident pump power of \SI{52}{mW}. Correcting for the imperfect mode matching of the pump beam, we infer an internal conversion efficiency greater than \SI{94}{\%}.

Far above the pump power of optimal conversion of \SI{50}{mW}, i.e.\ above \SI{600}{mW}, the output power at \SI{2128}{nm} was higher than predicted by the fit in Fig.~\ref{fig:conversion_efficiency}.
We assume that this was due to imperfect constructive interference of up-converted (re-converted) light with the pumped cavity mode, but operation at these powers should in any case be avoided to retain a stable and well controllable system.

\subsection{Interference of converted beams}
\label{sec:interference}

\begin{figure}
    \resizebox{\linewidth}{!}{\input{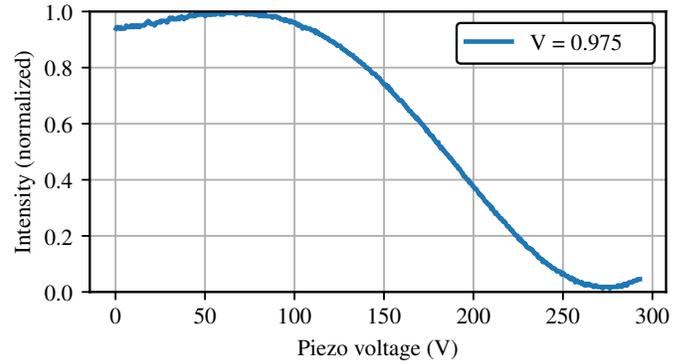}}
    \caption{
        Demonstration of spatial mode quality and frequency stability.
        Shown is the interference fringe \visibility\ generated by overlapping two independently generated \SI{2128}{nm} beams on a 50/50 beam splitter.
        The phase difference was scanned by applying a high voltage ramp to a piezo-mounted mirror.
        The achievable phase difference was limited by the travel range of the piezo, compared to the long wavelength.
    }
    \label{fig:interference}
\end{figure}

In the next step, we duplicated the DOPO setup and measured the visibility between the two independently created light fields at \SI{2128}{nm}.
For this, the beams were overlapped on a 50/50 beam splitter.
A piezo-mounted mirror allowed us to scan the relative phase between the two fields, cf.\ figure~\ref{fig:lasersystem}.
The resulting interference fringes were monitored on a photodiode in one of the beam splitter output ports, see figure~\ref{fig:interference}.
We observed a stable and stationary interference pattern, indicating quasi-identical operation of the two DOPOs and high coherence between the two fields.
A maximum interferometric visibility
\[
    V = \frac{I_\mathrm{max} - I_\mathrm{min}}{I_\mathrm{max} + I_\mathrm{min}} = 0.975
\]
was achieved, where $I_{\rm min}$ and $I_{\rm max}$ are the light powers detected at the minimum and maximum of the interference fringe, respectively.

The visibility is a measure for the coherence properties of light.
Deviations from $V=1$ are furthermore caused by non-perfect beam overlap, differing beam powers, as well as phase and amplitude fluctuations
\cite{sveltoPrinciplesLasers2010}.  
Our measured visibility is the sum of all these imperfections, which results in an upper limit for the coherence properties of the converted fields.
We assume that our visibility value was limited by imperfect beam overlap and even higher values would be achievable with careful alignment.
We conclude that the two individual wavelength conversions preserve coherence properties to a high degree and therefore that the conversion approach is suitable for interferometric applications.

\subsection{Power stability}

\begin{figure}
    \resizebox{\linewidth}{!}{\input{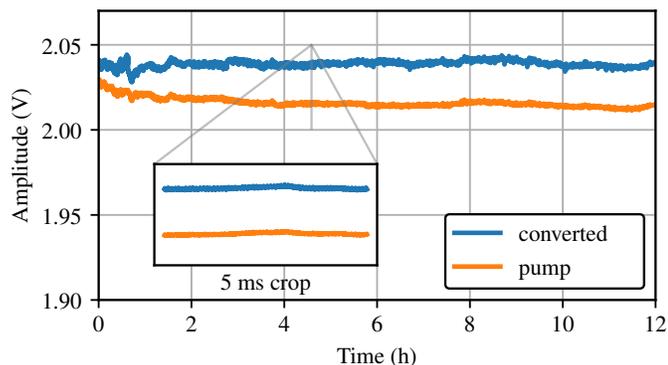}}
    \caption{
        Long and short term stability of the pumped and converted light.
        The first two hours show larger disturbances, which reduce after the optical setup has thermalized and traffic around the lab area reduced during the night.
        Note the scale on the y-axis.
    }
    \label{fig:stability}
\end{figure}

Converting the wavelength should retain the stability properties of the pump beam, without introducing additional power drifts or amplitude noise.
We measured the intensity of the pumped and converted light with separate photodiodes for twelve hours at a sampling rate of \SI{200}{kHz}.
Figure~\ref{fig:stability} shows the long and short term intensity stability, over the full time span and a zoom into a \SI{5}{ms} segment.
Both traces show small disturbances which are mostly correlated between both beams.
The relative error margins, starting from a settling time of 3 hours, are \SI{0.11}\% (pump) and \SI{0.15}\% (converted).
The overall stability of the \SI{2}{\micro m} light is thus similar to the stability of the NPRO pump laser.

\section{Conclusion}

We have reported on a stable, coherent light source at \SI{2128}{nm} using high-efficiency degenerate optical-parametric oscillation.
We showed long-term stability over many hours and proved strong coherence between two individually converted fields, with a measured visibility of $V=0.975$.

While our setup was optimized for low pump powers, our obtained value for the non-linearity allows for the conversion maximum to be adapted to arbitrary powers, by changing the reflectivities of the DOPO mirrors.
Our approach should thus enable the conversion of the highly stable \SI{1064}{nm} lasers used in current \gsw{} detectors with powers above \SI{100}{W} with external conversion efficiencies greater \SI{80}\%.
Furthermore, our scheme can be readily extended to create squeezed states of light, which have become an integral part of all current and future \gsw{} detectors.

\bigskip
\noindent\textbf{Funding Information.}
This research has been funded by the Deutsche Forschungsgemeinschaft (DFG, German Research Foundation) -- 388405737.
This article has LIGO document number P2000265 and Virgo codifier VIR-0668A-20.

\bigskip
\noindent\textbf{Disclosures.} The authors declare no conflicts of interest.

\bibliography{references}

\begin{thebibliography}{10}
\newcommand{\enquote}[1]{``#1''}

\bibitem{abbottObservationGravitationalWaves2016}
B.~P. Abbott and {et al.}, \enquote{Observation of {{Gravitational Waves}} from
  a {{Binary Black Hole Merger}},} {\protect\JournalTitle{Physical Review
  Letters}} \textbf{116} (2016).

\bibitem{ligoscientificcollaborationandvirgocollaborationGW170817ObservationGravitational2017}
{LIGO Scientific Collaboration and Virgo Collaboration}, \enquote{{{GW170817}}:
  {{Observation}} of {{Gravitational Waves}} from a {{Binary Neutron Star
  Inspiral}},} {\protect\JournalTitle{Physical Review Letters}} \textbf{119},
  161101 (2017).

\bibitem{abbottGW190814GravitationalWaves2020}
{LIGO Scientific Collaboration and Virgo Collaboration}, \enquote{{{GW190814}}:
  {{Gravitational Waves}} from the {{Coalescence}} of a 23 {{Solar Mass Black
  Hole}} with a 2.6 {{Solar Mass Compact Object}},} {\protect\JournalTitle{The
  Astrophysical Journal}} \textbf{896}, L44 (2020).

\bibitem{shoemakerGravitationalWaveAstronomy2020s2019}
D.~Shoemaker, \enquote{Gravitational-{{Wave Astronomy}} in the 2020s and
  {{Beyond}}: {{A View Across}} the {{Gravitational Wave Spectrum}},} Tech.
  rep. (2019).

\bibitem{levinInternalThermalNoise1998}
Y.~Levin, \enquote{Internal thermal noise in the {{LIGO}} test masses: {{A}}
  direct approach,} {\protect\JournalTitle{Physical Review D}} \textbf{57},
  659--663 (1998).

\bibitem{harryThermalNoiseInterferometric2002a}
G.~M. Harry, A.~M. Gretarsson, P.~R. Saulson, S.~E. Kittelberger, S.~D. Penn,
  W.~J. Startin, S.~Rowan, M.~M. Fejer, D.~R.~M. Crooks, G.~Cagnoli, J.~Hough,
  and N.~Nakagawa, \enquote{Thermal noise in interferometric gravitational wave
  detectors due to dielectric optical coatings,}
  {\protect\JournalTitle{Classical and Quantum Gravity}} \textbf{19}, 897--917
  (2002).

\bibitem{barsottiUpdatedAdvancedLIGO2018}
L.~Barsotti, S.~Gras, M.~Evans, and P.~Fritschel, \enquote{The updated
  {{Advanced LIGO}} design curve,} Tech. rep. (2018).

\bibitem{hildXylophoneConfigurationThirdgeneration2010}
S.~Hild, S.~Chelkowski, A.~Freise, J.~Franc, N.~Morgado, R.~Flaminio, and
  R.~DeSalvo, \enquote{A xylophone configuration for a third-generation
  gravitational wave detector,} {\protect\JournalTitle{Classical and Quantum
  Gravity}} \textbf{27}, 015003 (2010).

\bibitem{etscienceteamEinsteinGravitationalWave2011}
{ET Science Team}, \enquote{Einstein gravitational wave {{Telescope}}
  conceptual design study,}  (2011).

\bibitem{murrayIonbeamSputteredAmorphous2015}
P.~G. Murray, I.~W. Martin, K.~Craig, J.~Hough, R.~Robie, S.~Rowan, M.~R.
  Abernathy, T.~Pershing, and S.~Penn, \enquote{Ion-beam sputtered amorphous
  silicon films for cryogenic precision measurement systems,}
  {\protect\JournalTitle{Physical Review D}} \textbf{92}, 062001 (2015).

\bibitem{steinlechnerSiliconBasedOpticalMirror2018a}
J.~Steinlechner, I.~W. Martin, A.~S. Bell, J.~Hough, M.~Fletcher, P.~G. Murray,
  R.~Robie, S.~Rowan, and R.~Schnabel, \enquote{Silicon-{{Based Optical Mirror
  Coatings}} for {{Ultrahigh Precision Metrology}} and {{Sensing}},}
  {\protect\JournalTitle{Physical Review Letters}} \textbf{120}, 263602 (2018).

\bibitem{ligoscientificcollaborationInstrumentScienceWhite2019}
{LIGO Scientific Collaboration}, \enquote{Instrument {{Science White Paper}}
  2019,} Tech. rep., LIGO Scientific Collaboration (2019).

\bibitem{greenOpticalPropertiesIntrinsic1995}
M.~A. Green and M.~J. Keevers, \enquote{Optical properties of intrinsic silicon
  at 300 {{K}},} {\protect\JournalTitle{Progress in Photovoltaics: Research and
  Applications}} \textbf{3}, 189--192 (1995).

\bibitem{ETpathfinderDesignReport2020}
{ETpathfinder Team}, \enquote{{{ETpathfinder Design Report}},}  (2020).

\bibitem{Schnabel2010}
R.~Schnabel, N.~Mavalvala, D.~E. McClelland, and P.~K. Lam, \enquote{{Quantum
  metrology for gravitational wave astronomy.}} {\protect\JournalTitle{Nature
  communications}} \textbf{1}, 121 (2010).

\bibitem{LSC2011}
J.~Abadie, E.~al., and {The LIGO Scientific Collaboration}, \enquote{{A
  gravitational wave observatory operating beyond the quantum shot-noise
  limit},} {\protect\JournalTitle{Nature Physics}} \textbf{7}, 962--965 (2011).

\bibitem{Tse2019}
M.~Tse and {et al.}, \enquote{{Quantum-Enhanced Advanced LIGO Detectors in the
  Era of Gravitational-Wave Astronomy},} {\protect\JournalTitle{Physical Review
  Letters}} \textbf{123}, 231107 (2019).

\bibitem{Acernese2019}
F.~Acernese and {et al.}, \enquote{{Increasing the Astrophysical Reach of the
  Advanced Virgo Detector via the Application of Squeezed Vacuum States of
  Light},} {\protect\JournalTitle{Physical Review Letters}} \textbf{123},
  231108 (2019).

\bibitem{mansellObservationSqueezedLight2018}
G.~L. Mansell, T.~G. McRae, P.~A. Altin, M.~J. Yap, R.~L. Ward, B.~J.~J.
  Slagmolen, D.~A. Shaddock, and D.~E. McClelland, \enquote{Observation of
  {{Squeezed Light}} in the 2 {$M$}m {{Region}},}
  {\protect\JournalTitle{Physical Review Letters}} \textbf{120} (2018).

\bibitem{eckardtOpticalParametricOscillator1991a}
R.~C. Eckardt, C.~D. Nabors, W.~J. Kozlovsky, and R.~L. Byer, \enquote{Optical
  parametric oscillator frequency tuning and control,}
  {\protect\JournalTitle{Journal of the Optical Society of America B}}
  \textbf{8}, 646 (1991).

\bibitem{naborsCoherencePropertiesDoubly1990}
C.~D. Nabors, S.~T. Yang, T.~Day, and R.~L. Byer, \enquote{Coherence properties
  of a doubly resonant monolithic optical parametric oscillator,}
  {\protect\JournalTitle{JOSA B}} \textbf{7}, 815--820 (1990).

\bibitem{wongHighconversionefficiencyWidelytunableAllfiber2007a}
G.~K.~L. Wong, S.~G. Murdoch, R.~Leonhardt, J.~D. Harvey, and V.~Marie,
  \enquote{High-conversion-efficiency widely-tunable all-fiber optical
  parametric oscillator,} {\protect\JournalTitle{Optics Express}} \textbf{15},
  2947--2952 (2007).

\bibitem{arisholmOpticalParametricMaster2004}
G.~Arisholm, {\O}.~Nordseth, and G.~Rustad, \enquote{Optical parametric master
  oscillator and power amplifier for efficient conversion of high-energy pulses
  with high beam quality,} {\protect\JournalTitle{Optics Express}} \textbf{12},
  4189--4197 (2004).

\bibitem{marandiAllopticalQuantumRandom2012a}
A.~Marandi, N.~C. Leindecker, K.~L. Vodopyanov, and R.~L. Byer,
  \enquote{All-optical quantum random bit generation from intrinsically binary
  phase of parametric oscillators,} {\protect\JournalTitle{Optics Express}}
  \textbf{20}, 19322--19330 (2012).

\bibitem{kaneMonolithicUnidirectionalSinglemode1985}
T.~J. Kane and R.~L. Byer, \enquote{Monolithic, unidirectional single-mode
  {{Nd}}:{{YAG}} ring laser,} {\protect\JournalTitle{Optics Letters}}
  \textbf{10}, 65 (1985).

\bibitem{darsow-frommNQontrolOpensourcePlatform2020}
C.~{Darsow-Fromm}, L.~Dekant, S.~Grebien, M.~Schr{\"o}der, R.~Schnabel, and
  S.~Steinlechner, \enquote{{{NQontrol}}: {{An}} open-source platform for
  digital control-loops in quantum-optical experiments,}
  {\protect\JournalTitle{Review of Scientific Instruments}} \textbf{91}, 035114
  (2020).

\bibitem{zielinskaFullyresonantTunableMonolithic2017}
J.~A. Zieli{\'n}ska, A.~Zukauskas, C.~Canalias, M.~A. Noyan, and M.~W.
  Mitchell, \enquote{Fully-resonant, tunable, monolithic frequency conversion
  as a coherent {{UVA}} source,} {\protect\JournalTitle{Optics Express}}
  \textbf{25}, 1142--1150 (2017).

\bibitem{schonbeckaxelCompactSqueezedlightSource2018}
A.~Sch{\"o}nbeck, \enquote{Compact squeezed-light source at 1550 nm,} Ph.D.
  thesis, Universit\"at Hamburg, {Hamburg} (2018).

\bibitem{falkInstabilitiesDoublyResonant1971}
J.~Falk, \enquote{Instabilities in the doubly resonant parametric oscillator:
  {{A}} theoretical analysis,} {\protect\JournalTitle{IEEE Journal of Quantum
  Electronics}} \textbf{7}, 230--235 (1971).

\bibitem{smithThermoopticThermalExpansion2016b}
A.~V. Smith, J.~J. Smith, and B.~T. Do, \enquote{Thermo-optic and thermal
  expansion coefficients of {{RTP}} and {{KTP}} crystals over 300-350 {{K}},}
  {\protect\JournalTitle{arXiv:1607.03964 [physics]}}  (2016).

\bibitem{giordmaineTunableCoherentParametric1965}
J.~A. Giordmaine and R.~C. Miller, \enquote{Tunable {{Coherent Parametric
  Oscillation}} in {{LiNb O}} 3 at {{Optical Frequencies}},}
  {\protect\JournalTitle{Physical Review Letters}} \textbf{14}, 973--976
  (1965).

\bibitem{lastzkaNumericalModellingClassical2010}
N.~Lastzka, \enquote{Numerical modelling of classical and quantum effects in
  non-linear optical systems,} {{PhD Thesis}}, Universit\"at Hannover (2010).

\bibitem{sveltoPrinciplesLasers2010}
O.~Svelto and D.~C. Hanna, \emph{Principles of Lasers} ({Springer}, {New York},
  2010), 5th ed.

\end{thebibliography}

\bibliographyfullrefs{references}

\end{document}